\documentclass[12pt]{article}
\usepackage{graphicx}
\newcommand{\be}{\begin{equation}}
\newcommand{\ee}{\end{equation}}
\def\ben{\begin{equation}}
\def\een{\end{equation}}
\def\bea{\begin{eqnarray}}
\def\eea{\end{eqnarray}}
\def\pd{\partial}

\begin{document}
\pagestyle{plain}
\begin{titlepage}

\begin{flushright}
FIT-901\\
hep-th/99xxxxx
\end{flushright}
\vspace{15 mm}

\begin{center}
{\huge Born-Infeld strings between D-branes  }
\end{center}

\vspace{12 mm}

\begin{center}
K. Ghoroku ${}^a$ \footnote{gouroku@fit.ac.jp} 
and K. Kaneko ${}^b$  \footnote{kaneko@phys.kyusan-u.ac.jp} \\
\vspace{3mm}
${}^a$Department of Physics,
 Fukuoka Institute of Technology, \\
 Wajiro Higashi-ku 811-02, Fukuoka, 
 Japan\\
\vspace{3mm}
${}^b$Department of Physics,
Kyushu Sangyo University\\
Matsukadai, Higashi-ku, 813-8503, Fukuoka, 
Japan
\end{center}

\vspace{5 mm}
\begin{abstract}
\noindent

We examine the solutions of world-volume action for a D3-brane being
put near other D3-brane which is replaced
by the background configuration of bulk space.
It is shown that the BPS solutions are not affected by
the D3-brane background, and they are interpreted as dyonic strings
connecting two branes.
On the contrary, the non-BPS
configurations are largely influenced by the background D-brane, and 
we find that the solutions with pure electric charge 
cannot connect two branes. These solutions are corresponding to the bound
state of brane and anti-brane which has been found by Callan and Maldacena.

\end{abstract}
\end{titlepage}
\pagestyle{plain}
%\pagenumbering{roman}

%%%%%%%%%%%%%%%%%%%%%%%%%%%%%%%%%
\section{Introduction}

\vspace {0.5cm}
~~ Recently, classical
solutions of the Born-Infeld (BI) action, which is considered as 
the world-volume action of the D-brane, have been studied 
~\cite{callan,gibbons,hashi,bak}
in the flat background-space. They are classified into the BPS and
non-BPS solutions.
The stable BPS solutions are
interpreted as the strings which connect two D-branes 
separated by the infinite
distance. While the non-BPS solutions are not protected from the
dynamical fluctuations since they are not supersymmetric one.
Among them,
 non-BPS solutions with pure electric charge have attracted an attention
\cite{callan}.
It can be regarded as the half-part of the string which connects brane
and anti-brane with a finite distance to form a bound state of them.
The stability of such a non-BPS configuration
has been studied from a viewpoint
of the quantum mechanics \cite{callan,sav}.

It will be interesting to examine these solutions 
from a viewpoint
of the $SU(N)$ Yang-Mills theory which 
can be constructed by the stack of D3-branes. 
In this direction, some progress has been given by studying the 
world-volume action of test brane embedded in the background
of D-branes \cite{YM,CGS,CGST,gaun,CRS}.
The situation
of the symmetry breaking $SU(N)\to SU(N-1)\times U(1)$ is realized by
setting one of the branes far away from the others. Then
we can say that the classical solutions of the world-volume action
represent the soliton solutions such as monopoles or dyons 
appearing
in the non-Abelian Yang-Mills theory. 
It would be meaningful to study the non-BPS solutions
which are non-supersymmetric solution obtained
in the background of the type-IIB superstring theory, because 
they are expected to play some role
in the non-supersymmetric Yang-Mills theory.
 
The purpose of this paper is to study the classical solutions of 
the world-volume action 
in a situation that the test D-brane is set parallel to a background
D-brane(s). In the next section, the model is given. And in section
3.1, we show the BPS and non-BPS dyonic solutions 
in the flat background.
In sections 3.2 and 4, we solve the case of the D-brane background.
We find that the BPS solutions 
are not affected by the background, 
but the configuration of non-BPS solutions 
are influenced by the background. Especially, the pure electric solutions 
cannot reach at the position of the background brane(s). 
Conclusions are given in the final section.

%\vspace {1.0cm}

\section {D-brane action to be solved}

Even if we consider the supersymmetric case,
the fermionic coordinates are not necessary to obtain a classical
solution of the D-brane action. So we neglect them, and the bosonic
part of the effective action of
a D-p-brane being coupled to the background
can be written as follows,
\ben
S_{p+1}=-T_p\int d^{p+1}\xi ({\rm e}^{-\phi}
       \sqrt {- {\rm det} ( G_{\mu \nu} + F_{\mu \nu} )}
      +{1 \over (p+1)!}\epsilon^{i_1\cdots i_{p+1}}
      A_{i_1\cdots \cdots i_{p+1}})
\een
where $T_p$ denotes the tension of the D-p-brane and 
\ben
F_{\mu \nu}= \partial _\mu A _\nu - \partial _\nu A _\nu
\een
is the field strength of $U(1)$ gauge field living in the world volume. 
We also neglect here the antisymmetric tensor $B_{MN}$, which
should be added to $F_{\mu\nu}$, since the brane considering
here has no NS-NS charge. 
The metric $G_{\mu \nu}$ and the $(p+1)$-form are the pullback of 
the space-time metric $G_{MN}$ and the R-R $(p+1)$-form respectively,
\ben
G_{\mu \nu} =G_{MN}  \partial _\mu X^M \partial _\nu X^N 
\een
\ben
      A_{i_1\cdots \cdots i_{p+1}}
     = \pd_{i_1}X^{M_1}\cdots\pd_{i_{p+1}}X^{M_{p+1}}A_{M_1\cdots M_{p+1}}.
\een
The embedding of the world volume into the
target space is described by $X^M(\xi^\mu)$ as a function of the $p+1$ 
dimensional world-volume coordinates $\xi^\mu$.

Our purpose is to solve the equations of motion for the above action
$S_{p+1}$ under the target space configuration for $\Phi, G_{MN}$ and 
$A_{M_1\cdots M_{p+1}}$, which are obtained as the D-p-brane configuration
by solving the ten-dimensional supergravitational theory. Before solving
the equations of D-brane action in this way, 
we briefly review the D-brane configurations.
%%%%%%%%%%%%%%%%%%%%%%%%%%%%%%%%%%%%%%%%%%%%%%%%%%%%%%%%

They are obtained by solving
the supergravity effective action corresponding to the 
superstring theory \cite{Duff,Horowitz}.
The bosonic part of the low energy effective action in 
$d$-dimensional \footnote{ Here $d$ is used to denote the target space
dimension, but we consider the case of $d=10$ throughout this paper}
target space can be
written as
\ben
S_{\rm d}=-\int d^{d}x \sqrt{-g}\left\{{\rm e}^{2\Phi}\left(R+4(\pd\Phi)^2
           -{1\over 12}H^2_3\right)
           -{1 \over 2(p+2)!}F_{p+2}^2 \right\}
\een
where $H_3(=dB)$ is the three form
field strength of the antisymmetric tensor $B_{MN}$ and
$F_{p+2}(=dA_{p+1})$ denotes the $(p+2)$-form field-strength of the 
$(p+1)$-form
R-R potential $A_{p+1}$.

In solving the classical equations of $S_{d}$
with respect to $g_{MN}$, $\Phi$
and $A_{p+1}$, 
several ansatzs are adopted. 
The $d$-dimensional
coordinates $x^M$ are denoted by separating them into the tangential
($x^{\mu}$) and the transverse part ($y^m$) to the D-p-brane
as $x^M=(x^{\mu},y^m)$, where $\mu=0\sim p$ and
$m=p+1\sim d-1$. Then take the following ansatzs
\ben
 ds^2={\rm e}^{2A}\eta_{\mu\nu}dx^{\mu}dx^{\nu}
          +{\rm e}^{2B}\delta_{mn}dy^{m}dy^{n}~, \label{brane1}
\een  
\ben
 A_{p+1}=-{\rm e}^Cdx^0\wedge dx^1\wedge\cdots \wedge dx^p~, \label{brane2}
\een
where $A, B, C$ and $\Phi$ are the functions of $y$ only,
here $y=\sqrt{\delta_{mn}y^my^n}$. The last statement means that
the solutions are assumed to be invariant under $P_{p+1}\otimes SO(d-p-1)$ 
transformation. 
Then
$A, B, C$ and $\Phi$ are all expressed by a common function, $H(y)$,
which is determined by the field equations derived by $S_d$.
The final result is obtained as follows:
\ben
 {\rm e}^{2A(y)}=H(y)^{-1/2},~~~{\rm e}^{2B(y)}=H(y)^{1/2}~,~\label{eq:dsol}
\een
\ben
 {\rm e}^{C(y)}=H(y)^{-1}-1, ~~~~~{\rm e}^{2\Phi}=H(y)^{(3-p)/2},  \label{brane3}
\een
where
\ben
H(y) = 1 + 2Q \kappa T_p G(y) ~~,
\label{sol}
\een
\ben
G(y) = \left\{ \begin{array}{cc}
 \left[ \tilde{d} | y|^{\tilde{d}}
\Omega_{\tilde{d}+1}\right]^{-1} &  ~~~ \tilde{d}>0 ~~,\\
  - \frac{1}{2 \pi} \log |y|  &   ~~~\tilde{d}= 0 ~~.
\end{array} \right\}~,
\label{solu}
\een
and $\tilde{d}=d-p-3$. Here 
$\Omega_{q}={ 2 \pi^{(q+1)/2}}/{ \Gamma \Big((q+1)/2\Big)}$
denotes the area of a unit $q$-dimensional sphere $S^q$.

Here $B_{MN}=0$ since the D-brane considered above has 
one kind of the R-R charge and no NS-NS charge, and
we use the form of $p=3$ in the following.

\vskip 1.5cm

%%%%%%%%%%%%%%%%%%%%%%%%%%%%%%%%%%%%%%%%%%%%%%%%%%%%%%%
\section{Solutions of D-brane action}

Here we discuss on the classical solutions of $S_{p+1}$ in the background 
D-brane(s) which is set parallel to the test brane 
considering now.
The classical equations of $S_{p+1}$ are solved for $p=3$ by adopting
the static gauge for the diffeomorphism invariance,
for which the 
world-volume coordinates $\xi^{\mu}$
are equated with the $p+1$ spacetime coordinates as 
\ben
X^M=\xi^\mu, ~~~~~~M=0,1,\dots ,p.
\een
The remaining coordinates, $X^m$, are treated as scalar fields in the
brane world-volume, but we retain only one coordinate among them as
a field which is responsible for the configurations of the brane and
it is denoted here by $X(\xi)$. This coordinate is taken in the direction
perpendicular to the other brane.

Then we can write $S_{p+1}$ with $p=3$ as follows,
\ben
S_4=-T_3\int d^4\xi\left( {1\over H}(\sqrt{D}-1)\right ) \, \label{d3br}
\een
where the field-independent constant
is subtracted from the above lagrangian density
for the brevity, and 
\bea
&& D=(1-H\vec{E}^2)(1+H(\nabla X)^2)+H^2(\vec{E}\cdot \nabla X)^2
 -H\dot{X}^2+2H^2\vec{E}\cdot (\vec{B}\times\nabla X)\dot{X} \nonumber \\
&&~~~~+ H\vec{B}^2(1-H\dot{X}^2)+H^2(\vec{B}\cdot\nabla X)^2
   -H^2(\vec{B}\cdot\vec{E})^2  ,\nonumber \\ 
\eea
\ben
H= 1 + 2Q \kappa T_3 G(X_m-X) ~~,
\een
where $X_m$ denotes the distance between two branes.
The U(1) gauge fields are denoted by the conventional,
electromagnetic fields $\vec{E}$ and $\vec{B}$.
%%%%%%%%%%%%%%%%%%%%%%%%%%%%%%%%%%%%

\subsection{Solutions in the flat background}

~~Firstly, we solve the equations for $H=1$: the case of the infinite
distance between the two branes. The solutions are restricted to the
static one. In this case, $D$ is written as,
\ben
 D = (1-\vec{E}^{2})(1+({\nabla}X)^{2})+(\vec{E}\cdot{\nabla}X)^{2}
     +\vec{B}^{2}+(\vec{B}\cdot{\nabla}X)^{2}-(\vec{B}\cdot\vec{E})^{2},
              \label{eq:8}
\een
and the following equations are obtained,
\begin{eqnarray}
 {\nabla}\cdot\{ [(1-\vec{E}^{2}){\nabla}X
      +(\vec{E}\cdot{\nabla}X)\vec{E}+
     (\vec{B}\cdot{\nabla}X)\vec{B} ]/\sqrt{D} \}=0, 
                       \hspace{1cm} \label{eq:5} \\
 {\nabla}\cdot\{ [(1+({\nabla}X)^{2})\vec{E}-
         (\vec{E}\cdot{\nabla}X){\nabla}X+
          (\vec{B}\cdot\vec{E})\vec{B} ]/\sqrt{D} \} = 0, 
                       \hspace{1cm} \label{eq:6} \\
 {\nabla}\times\{ [\vec{B}+
           (\vec{B}\cdot\nabla X)\nabla X
          -(\vec{B}\cdot\vec{E})\vec{E}) ]/\sqrt{D} \}=0. 
                  \hspace{1cm} \label{eq:7} 
\end{eqnarray}

Since we are interested in isotropic solutions, we take the 
following ansatz:
\ben
  \vec{E}=f(r)\hat{r}~,~~~\vec{B}=g(r)\hat{r}~,~~~\nabla X=X'(r)\hat{r}\, ,
          \label{ans1}
\een
where $\hat{r}$ represents the unit vector in the radial direction, and
a prime denotes a differentiation with respect to $r$.
Then, Eqs. (\ref{eq:5}) and (\ref{eq:6}) are expressed as
\begin{eqnarray}
{\nabla}\cdot \{ X'\sqrt{\frac{1+g^{2}}{1-f^{2}+X'^{2}}}\hat{r} \}=0, \label{eq:15} \\
{\nabla}\cdot \{ f\sqrt{\frac{1+g^{2}}{1-f^{2}+X'^{2}}}\hat{r} \}=0. \label{eq:16} 
\end{eqnarray}
From Eqs. (\ref{eq:15}) and (\ref{eq:16}), the function $f$ is proportional to $X'$
\begin{eqnarray}
 f = \alpha X', \label{eq:17}
\end{eqnarray}
where $\alpha$ is a constant. Since the ansatz (\ref{ans1})
satisfies automatically Eq. (\ref{eq:7}) for any $g$, we cannot determine the function $g$ since the number of the independent equations is
reduced to two. So we need further ansatz or assumption to obtain the solution.
We now
assume that the magnetic field $\vec{B}$ is given by some appropriate form
as an arbitrary external field, then we can solve the equation 
Eq. (\ref{eq:15}) or (\ref{eq:16}) as follows,
\begin{eqnarray}
 X(r) = \int_{r}^{\infty}dr\frac{A}{\sqrt{(1+g^{2})r^{4}-r_{0}^{4}}}, \label{eq:18}
\end{eqnarray}
where $A$ denotes an integral constant and 
$r_{0}^{4}=(1-\alpha^{2})A^{2}$. Here we consider the 
purely electric ($g=0$) case, namely without the external magnetic field, then
the solution (\ref{eq:18}) reduces to the BPS state for $\alpha=1$ and to the 
non-BPS solution given in Ref.\cite{callan} for $\alpha\neq 1$.

%===  Fig.  ========================================================
\begin{figure}
\begin{center}
   \includegraphics[width=7cm,height=7cm]{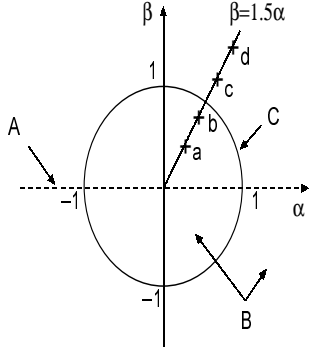}
 \caption{The static solutions are classified on $\alpha - \beta$ plane. 
          They are represented by three types of solutions: 
          (A)The pure electric solution, (B)the dyonic solution, 
          and (C)the BPS solution.  
          The symbols of plus denote the points on the line
$\beta=1.5\alpha$
           for (a) $\alpha=0.3$, (b) $0.5$, (c) $0.7$ and (d) $0.9$. }
\end{center}
\end{figure}
%====================================================================

Next, consider the case adding one more ansatz to Eq. (\ref{ans1})
as follows:
\begin{eqnarray}
 g = \beta X', \label{eq:19}
\end{eqnarray}
where $\beta(\neq 0)$ is a constant. In this case, the string is considered as
the source of both the electric and the magnetic charges, so we call this type
as the dyonic string. The equation is solved as
\begin{eqnarray}
 X(r) = \int_{r}^{\infty}dr \biggl{[} \frac{ -\{ 1-(1-\alpha^{2})\frac{A^{2}}{r^{4}} \} + \sqrt{(1-(1-\alpha^{2})\frac{A^{2}}{r^{4}})^{2}+4\beta^{2}\frac{A^{2}}{r^{4}}} }{2\beta^{2}} \biggr{]}^{1/2}. \hspace{1cm} \label{eq:21}
\end{eqnarray}
For $\alpha^{2}+\beta^{2}=1$, this solution reduces to the BPS dyonic solution,
\begin{eqnarray}
 X(r) = \frac{A}{r}. \label{eq:20}
\end{eqnarray}
In this case,
the Bogomol'nyi equations are satisfied by the saturation of the bound
\begin{eqnarray}
 E \geq \alpha (\vec{E}\cdot \vec{\nabla}X)+\beta (\vec{B}\cdot \vec{\nabla}X). \label{eq:23}
\end{eqnarray}
The solutions obtained here are classified in the parameter
space $\alpha$ and $\beta$ as shown in Fig. 1.

\begin{figure}[htbp]
\includegraphics[width=10cm,height=7cm]{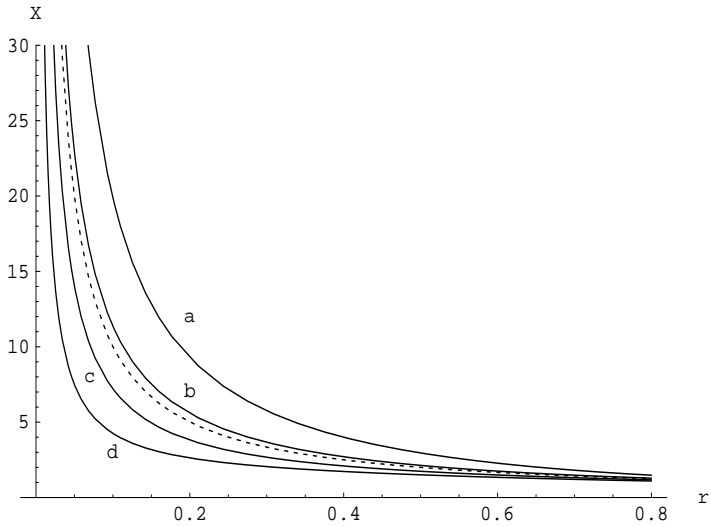}
\caption{\footnotesize{The typical dyonic non-BPS solutions, for 
$\beta=1.5\alpha, A=1$. And the curves represent 
for (a) $\alpha=0.3$, (b) $0.5$, (c) $0.7$ and (d) $0.9$ in Fig. 1.
The boundary conditions are taken as
$X(4)=A/4$ for all curves. }}
\end{figure}

There are many number of non-BPS dyonic string solutions, and some typical
one are shown in the Fig.2 along the line in the parameter space shown in
Fig.1.

%%%%%%%%%%%%%%%%%%%%%%%%%%%%%%%%%%%%%
\subsection{Solutions in the background D-brane(s)}

Here we consider the solutions for the case of $H\neq 1$. In this case,
the Wess-Zumino term in the action $S_4$ given in Eq. (\ref{d3br})
cannot be neglected as in 
the previous section. To see its important meaning, 
we notice that $S_4$ vanishes for
$\vec{E}=\vec{B}=0$. This is a result of the cancellation between
the BI and the WZ terms, and it
reflects that there is no force between two 
parallel branes \cite{polchin}. We therefore expect that the WZ term
would play an important role in solving the equations for $H\neq 1$.

Here we solve the equations of motion of $S_4$ in the background
configuration of the parallel D3-brane(s), so the force is absent
for the trivial solution of $\vec{E}=\vec{B}=0$. But it is expected that
the non-trivial solutions, which are obtained in the background of
flat space-time by solving the BI-action, would be affected by the background
configuration. The surviving solutions should be deformed and restricted.

The equations are written as follows:
\ben
   {\partial H \over\partial X}F_1+\nabla\cdot\vec{F_2}=0 \, , \label{eqX}
\een
\ben
 \nabla\cdot {1\over \sqrt{D}}\left(
    (1+H(\nabla{X})^2)\vec{E}
     +H[(\vec{B}\cdot\vec{E})\vec{B}-(\vec{E}\cdot \nabla X)\nabla X]
       \right)=0 \, , \label{eqE}
\een
\ben
 \nabla\times {1\over \sqrt{D}}\left(
    (\vec{B}
     +H[(\vec{B}\cdot \nabla X)\nabla X-(\vec{B}\cdot\vec{E})\vec{E}]
       \right)=0 \, , \label{eqB}
\een
where ${\partial H \over\partial X}=-4\tilde{Q}/(X_m-X)^5$ and
\bea
&& F_1={1\over H^2}(\sqrt{D}-1)+{1\over 2H\sqrt{D}}\Biggl(
    \vec{E}^2(1+H(\nabla{X})^2)-(\nabla X)^2(1-H\vec{E}^2)
                \nonumber \\
&&~~-2H[(\vec{B}\cdot\nabla X)^2+(\vec{E}\cdot\nabla X)^2
     -(\vec{B}\cdot\vec{E})^2]-\vec{B}^2
        \Biggr)  , \label{eqX1} 
\eea
\ben
     F_2= {1\over \sqrt{D}}\left(
    (1-H\vec{E}^2)\nabla X
     +H[(\vec{E}\cdot\nabla X)\vec{E}+(\vec{B}\cdot \nabla X)\vec{B}]
       \right) \, . \label{eqX2}
\een
Here the time-dependence of the fields are neglected since the
solutions are restricted to the static one. Obviously
the effect of the background configuration appears through $H$, and
Eqs. (\ref{eqX}) $\sim$ (\ref{eqB}) are respectively
reduced to Eqs. (\ref{eq:5})
$\sim$ (\ref{eq:7}) for $H=1$. The most characteristic
feature is seen in the first term of the l.h.s. of Eq. (\ref{eqX}). 
And the WZ term
is also included in $F_1$ of this term.
For $\tilde{Q}=0$, this term disappears, and the 
second term reduces to Eq. (\ref{eq:5})
given in the flat background. The first term therefore represents
a newly appeared constraint due to the non-trivial background 
configuration.
Due to this term, the above equations can be solved with ansatzs
(\ref{ans1}) and (\ref{eq:17}) only since the new constraint determines the 
magnetic field $g(r)$ as seen below.
    
We solve the above equations by taking the ansatz (\ref{ans1}) used in 
the previous subsection. Then
the above equations are rewritten as follows:
\ben
    {\partial H \over\partial X}F_1+
 \nabla\cdot\left( {X'\over\sqrt{D}}(1+Hg^2)\hat{r}\right)=0~, \label{eqX22}
\een
\ben
 \nabla\cdot\left( {f\over\sqrt{D}}(1+Hg^2)\hat{r}\right)=0~, \label{eqE2}
\een
\ben
 \nabla\times\left( {f\over\sqrt{D}}(1+H(X'^2-f^2))\hat{r}\right)=0~, 
\label{eqB2}
\een
where
\ben
    F_1= {1\over 2H^2\sqrt{D}}\left( 
      2(D-\sqrt{D})+H(f^2-X'^2-g^2)+2H^2g^2(f^2-X'^2)\right)~, 
\label{eqX12}
\een
\ben
        D=(1+Hg^2)(1+H(X'^2-f^2))~. \label{det}
\een
With the same way as in the previous subsection,
the number of equations is reduced to two since
 Eq. (\ref{eqB2}) is satisfied by any functional form of
$f$, $g$ and $X$. 

Then we need one more ansatz
which gives one relation among these three functions. To find
similar solutions to the one obtained in the previous subsection,
we take the following ansatz
\ben
 f=\alpha X' ~. \label{ansatz2}
\een
It should be noticed that this was not an ansatz but a solution of
the equations in the previous subsection.
Under this ansatz, Eq. (\ref{eqE2}) is solved as
\ben
  {f\over\sqrt{D}}(1+Hg^2)={A_2\over r^2}~, 
   \label{sol1}
\een
where $A_2$ is an integral constant. The Eq. (\ref{eqX22})
is rewritten as
\ben
  F_1=0~,  \label{eqX3}
\een 
since the second term of this equation
vanishes, $i.e.$ $\nabla F_2=0$.
We notice that the above equation (\ref{eqX3})
does not appear in the case of the flat
background. As a result, the magnetic field $g(r)$ can not be determined.
In the present case, we can obtain the following solution
due to Eq. (\ref{eqX3}), 
\ben
  X={A_2\over r}~,~~f=-\alpha {A_2\over r^2}~,
~~g=-\sqrt{1-\alpha^2}{A_2\over r^2}~. \label{sol2}
\een
This solution represents the dyonic string with both electric and magnetic
charges, and the BPS bound is saturated by this solution since it is
represented as
\ben
  \vec{E}=\cos(\theta)\nabla X~, ~~~\vec{B}=\sin(\theta)\nabla X~,
\een
by the parametrization
$\cos\theta=\alpha(\leq 1)$. Here, we notice the following points:
(i) The ansatz (\ref{ansatz2}), which yields a non-BPS 
solution in the case of the flat background, leads to the BPS solution due
to the Eq. (\ref{eqX3}). (ii) The form of the BPS solution obtained is
independent on the parameter of the background since it is not deformed by
the background configurations.
(iii) The effect of the background configuration appears indirectly
 through the induced world-volume metric in the brane action. Namely,
$r$ has a minimum value, $r_0=A_2/X_m$ \cite{gaun}, where
the solution touches the opposite brane(s) since $X(r_0)=X_m$.
At this point, the proper distance in the world-volume of the
brane becomes infinite since the
induced metric diverges. Then we arrive at the configuration of the 
dyonic string which connects two branes with a finite distance.

From the fact (iii),
the energy of the string part is obtained as a finite value 
by integrating the energy density of the system in the range $r_0<r<\infty$ as
follows,
\ben
 E= 4\pi A_2T_3X_m ~,
\een
which represent the energy of the string of the length $X_m$ with the tension
of $4\pi A_2T_3$. It is expected that
this object with a finite energy would appear as a dyon
in the Yang-Mills theory.

\section{Non-BPS solutions}

We now turn to the non-BPS solutions. 
As seen in the subsection 4.1, the non-BPS solutions obtained
in the flat space
are separated into two groups from their property. One is the pure 
electric string which connects to the anti-brane to form a bound state. 
The second is the dyonic string which could arrive at the other brane(s)
at the infinite distance. In the case of the finite distance also,
these two types of solutions can be found 
by taking other ansatz than the one given in the previous section,
(\ref{ansatz2}). 

\subsection{ Pure electronic case }

First we consider the case of the pure electric solution.
This is obtained by taking the following ansatz,
\ben
 g=0 .
\een
In this case, we can obtain the following
equation for $X$ from Eqs. (\ref{eqX22}) and (\ref{sol1}),
\ben
 X''+{2\over r}X'-{X'\over 2}(\ln D)'=\frac{2\tilde{Q}(\sqrt{D}-1)^2}
       {X^5H^2} \, , \label{eqX4}
\een
where
\ben
 D={1+HX'^2 \over 1+H(A_{2}/r^2)^2} \, .
\een
It is easily seen that Eq. (\ref{eqX4}) is reduced to 
the equation of the flat background in the 
limit of $\tilde{Q}=0$ $(H=1)$.
In fact, we obtain the following solution for $\tilde{Q}=0$,
\ben
 X'={c_1/r^2\over \sqrt{1-c_2/r^4}} \, ,
\een
where $c_1$ and $c_2$ are constants depending on the boundary 
conditions. This is equivalent to the solution (\ref{eq:18}).

While Eq. (\ref{eqX4}) for $H\neq 1$
is highly nonlinear, so it is difficult to solve it
analytically. Then we give the solutions numerically. 

\begin{figure}[htbp]
\includegraphics[width=14cm,height=8cm]{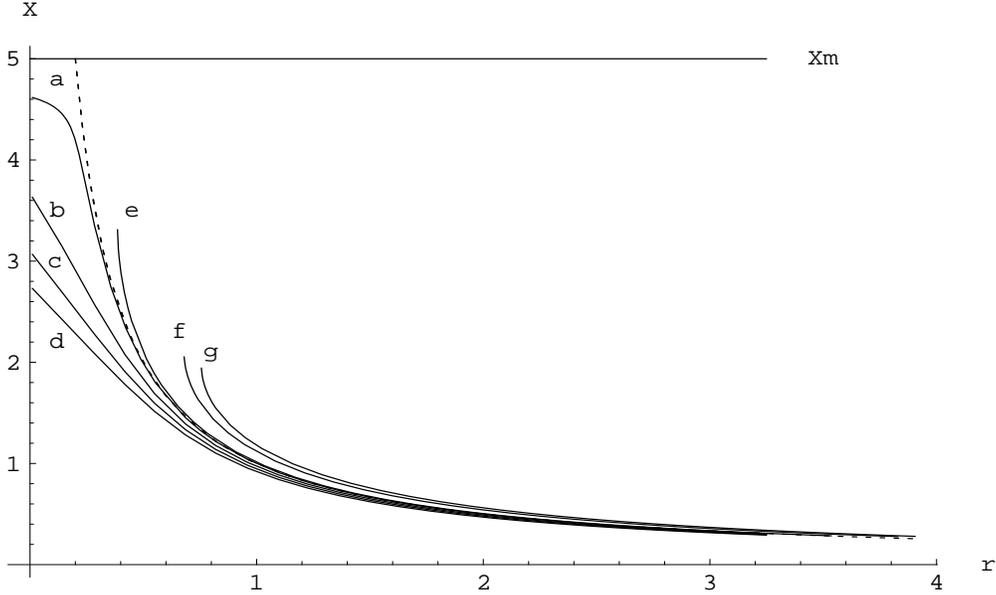}
\caption{\footnotesize{The typical two types of non-BPS solutions, for 
$\tilde{Q}=1, X_m=5, A_2=1$. 
The boundary conditions are taken as
$X(10)=0.1$ and (a) $X'(10)=-0.00974$, (b) $-0.00948$,
(c) $-0.00923$, (d) $-0.00898$ for the first type, and 
(e) $X'(10)=-0.0101$, (f) $-0.011$,
(g) $-0.0115$ for the second type. }}
\end{figure}

The resultant solutions are shown in Fig. 3,
where the BPS solution,
$X=A_{2}/r$, is also shown by the dotted line for the comparison.
In the Fig.3, two types of non-BPS solutions are seen.
They are obtained here by giving the boundary conditions at an appropriate
point ($r=r_B$) as
\ben
   X(r_B)={A_2\over r_B}~, ~~X'(r_B)=-\gamma {A_2\over r_B^2}~,
\een
where $\gamma$ is a parameter. The solutions of the first group $(a)\sim(d)$
are obtained for $\gamma <1$, and the second group $(e)\sim(g)$ 
are for $\gamma >1$.
We notice that the BPS solution is obtained for $\gamma =1$.
In this sense,
the BPS solution is the critical one which 
separates the two types of solutions.

The first type of solution covers all region of $r$, and
it ends at $r=0$ with finite value of $X(0)$ and negative value of
$X'(0)$. So the shape
of this configuration has a cusp at $r=0$ and the configuration is singular
at this point. 

The second type of solution is bounded as $r\geq r_1$, where $r_1$ varies
depending on the boundary conditions and $X'(r_1)=\infty$. Due to this
property, the 
solution can be
connected to the other half with the opposite orientation
\cite{callan} 
to make the bound state of brane and anti-brane.

These solutions have the same qualitative properties with the one
obtained in the flat background. However they are affected by the 
non-trivial background in the present case.
The most prominent influence is seen in the fact
that these solutions could not 
reach at the position of the other brane(s), $X=X_m$. 
This is understood as follows. From Eqs. (\ref{det}) and (\ref{sol1}), we
obtain
\ben
 f^2={H^{-1}+X'^2 \over H^{-1}(r^2/A_{2})^2+1} \, .
\een
Then we find the relation
$f=X'$ at $X=X_m$, where $H^{-1}=0$, and we obtain the
BPS solution from this condition. This implies that only the BPS solution
can arrive at $X=X_m$.
So there is no non-BPS electric
string state, which connect two parallel branes with a finite energy.
Such a object is restricted to the BPS saturated strings.

\begin{figure}[htbp]
\includegraphics[width=14cm,height=7cm]{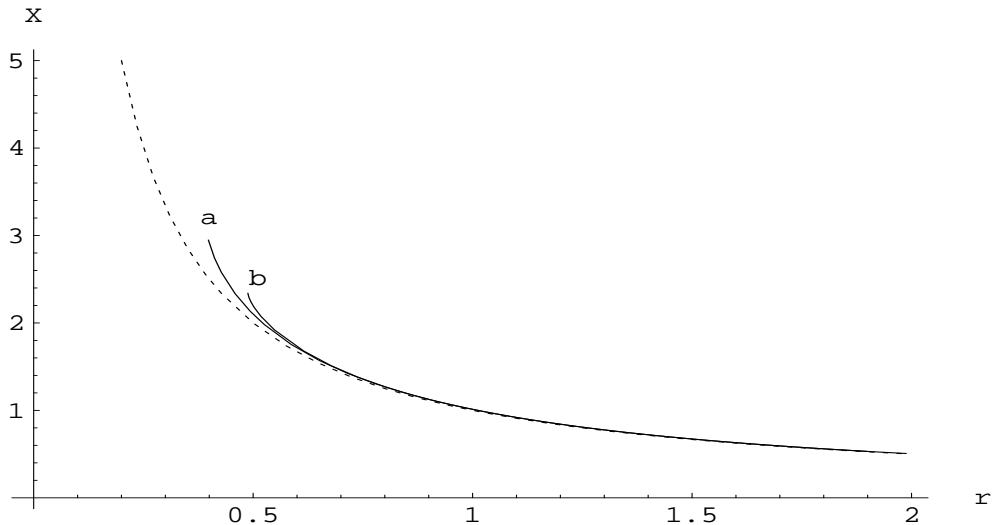}
\caption{\footnotesize{The typical non-BPS solutions
for $X_m=5, A_2=1$, (a) $\tilde{Q}=1$ and (b) $\tilde{Q}=70$
with the boundary condition $X(10)=0.1$ and $X'(10)=-0.0101$.
}}
\end{figure}

As for the $\tilde{Q}$ dependence of the solutions, we can see
the typical effect
from the solutions as shown in Fig. 4, where the second type of solutions
for two different $\tilde{Q}$ value are given. 
The $\tilde{Q}$ dependence is small, but it can be seen near the 
end point of the solution. The larger $\tilde{Q}$ becomes 
the wider the radius of the tube grows, and the end point of the 
string configuration goes back. This implies that the bound state of
the brane and anti-brane would be pushed to vanish through the
annihilation of them near the branes.

\subsection{ Dyonic case }

Next, we consider the dyonic non-BPS string configurations under the
D-brane background. The equation for this solution is obtained by taking
the ansatz (\ref{eq:19}), $g=\beta X' ,$
which is given in the section 3.1 to obtain the dyonic
solution in the flat background. In the present case, Eq. (\ref{eq:17})
is not used differently from the the case of section 3.1. 
In terms of this ansatz (\ref{eq:19})
and Eq. (\ref{sol1}), which is the solution of (\ref{eqE2}), we obtain
\ben
f^2={1+HX'^2\over 1+H[\beta^2X'^2+(A_2/r^2)^2]}({A_2\over r^2})^2~.
      \label{sol3}
\een
Substituting Eq. (\ref{sol3}) and (\ref{eq:19}) into Eq. (\ref{eqX22}), 
we can solve Eq. (\ref{eqX22}) with respect to $X(r)$ by rewriting
it as the differential equation of $X(r)$. 
We solve the differential equation numerically. 

We firstly discuss the dyonic solutions which
are corresponding to the solutions $(g)$ in Fig.3 at the limit
of $\beta=0$. Namely we take the same boundary conditions with the one of
$(g)$ in Fig.3 in solving the equation considered here for 
$\beta\neq 0$. In Fig.5, the solutions are
shown for various values of $\beta$. 
The $\beta$-dependence of the solutions is seen from the Fig.5, 
and the behaviors are similar to the case of the flat background. 

%%%%%%%%%%%%%%%%%%%%%%% Fig.5 %%%%%%%%%%%%%%%%%%%%%%%%%%%%%%%
\begin{figure}[htbp]
\includegraphics[width=14cm,height=7cm]{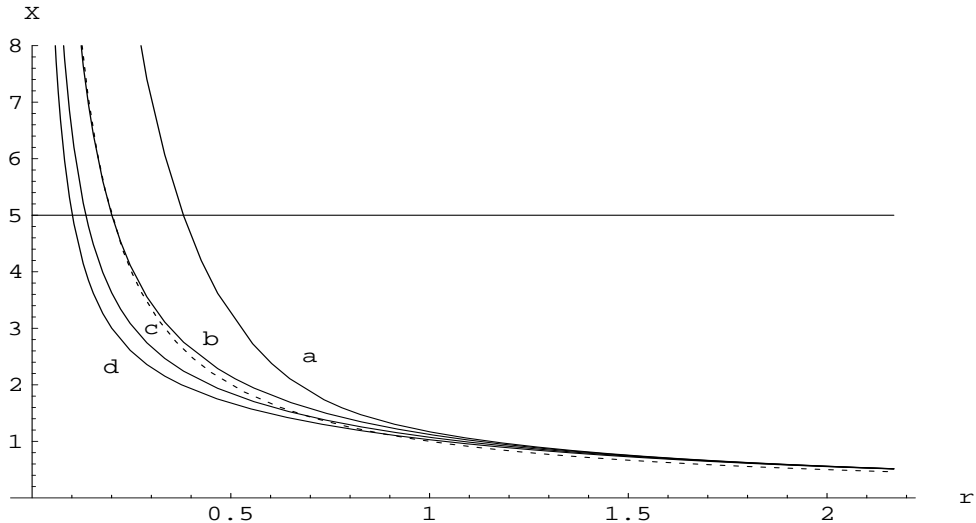}
\caption{\footnotesize{The typical dyonic non-BPS solutions
for $X_m=5, A_2=1$, (a) $\beta=0.2$, (b) $\beta=0.6$,
(a) $\beta=1.0$ and (b) $\beta=1.4$
with the boundary condition $X(10)=0.1$ and $X'(10)=-0.0115$.
}}
\end{figure}
%%%%%%%%%%%%%%%%%%%%%%%%%%%%%%%%%%%%%%%%%%%%%%%%%%%%%%%%%%%%

These solutions exceed $X_m$ which was the bound for the non-BPS solution
of $\beta=0$. 
This property is also seen in the case of the flat background, 
where this configuration extends to $X=\infty$ at $r=0$. But the shapes
of this configurations are affected by the D-brane background. This point
is different from the case of the flat background.
This is seen
from the results shown in Fig.6, in which 
$\tilde{Q}$-dependence for the solution is presented. 
%%%%%%%%%%%%%%%%%%%%%%% Fig.6 %%%%%%%%%%%%%%%%%%%%%%%%%%%%
\begin{figure}[htbp]
\includegraphics[width=7cm,height=7cm]{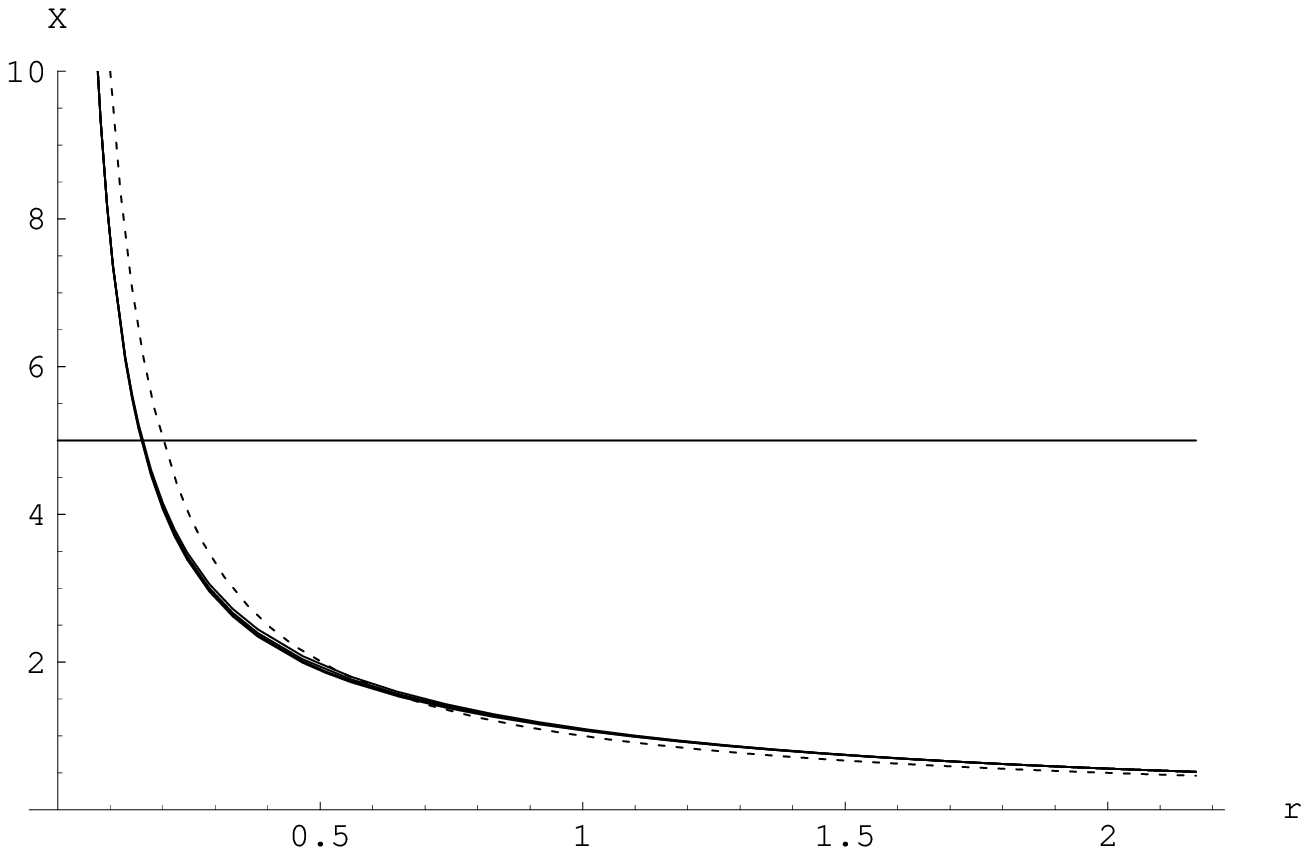}
\includegraphics[width=7cm,height=7cm]{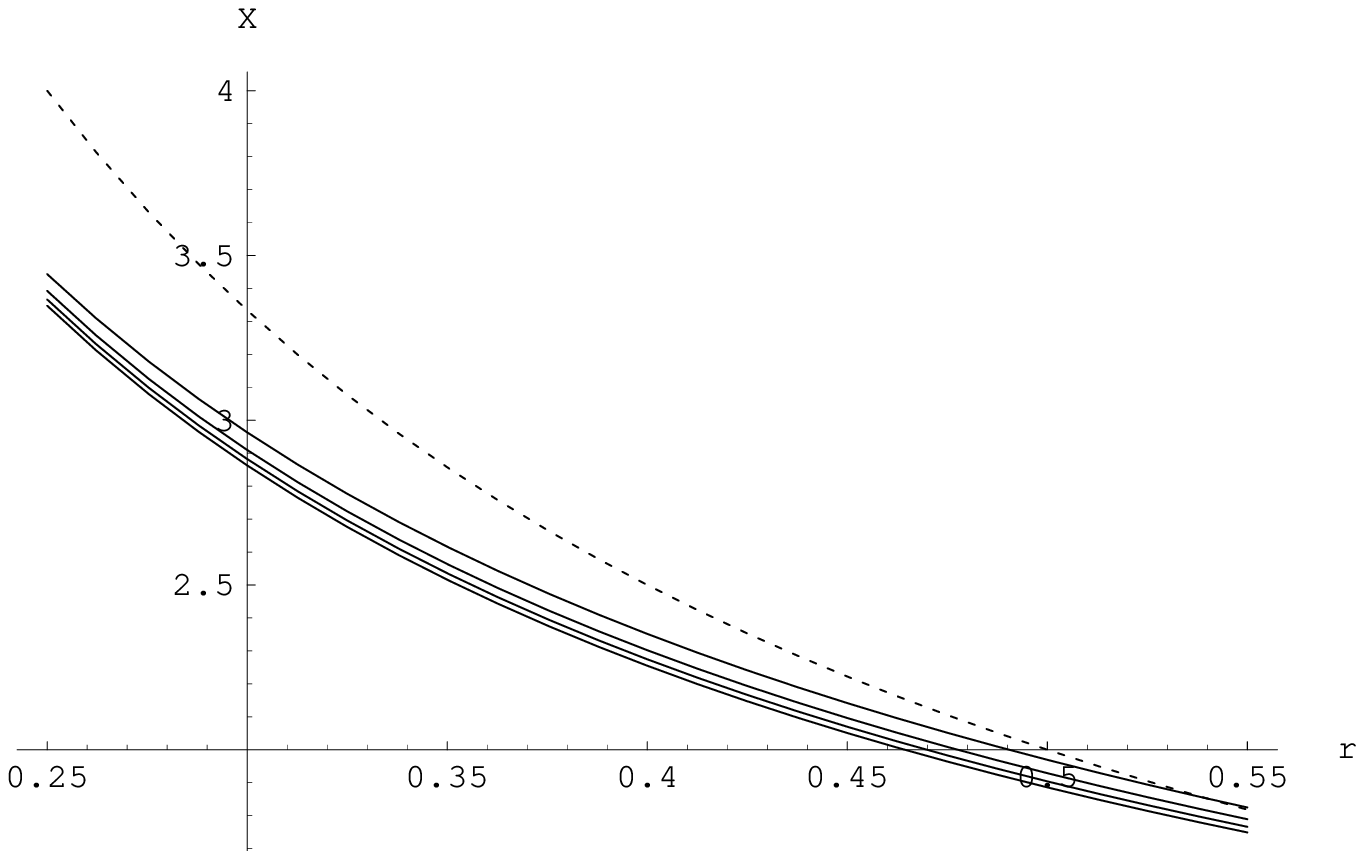}
\caption{\footnotesize{The $\tilde{Q}$-dependence of 
the dyonic non-BPS solutions
for $X_m=5, A_2=1$, $\beta=0.8$ with the boundary condition $X(10)=0.1$ 
and $X'(10)=-0.0115$. From the above curve, 
the curves for $\tilde{Q}=1, 101, 201, 301$ are
shown.
}}
\end{figure}
%%%%%%%%%%%%%%%%%%%%%%%% Fig.6 %%%%%%%%%%%%%%%%%%%%%%%%%%%
Although the $\tilde{Q}$-dependence is small, we can see it near
the neck ($0.2<r<6$) of the throat by extending the scale of $X(r)$. 
This $\tilde{Q}$-dependence is the main difference from the BPS
solutions which are independent on the background. 

We should notice the following fact.
For any solution, the
metric of the world-volume action is determined from the same D-brane
background, and it has a singularity at $X=X_m$. So the configuration
should be cut at this point, then we can say that the configuration
obtained here represents the dyonic string which connects two D-branes
with a finite distance. But these solutions are not the BPS states, so
some dynamical corrections would modify the configurations obtained here.
This is a dynamical problem, which would be related to the
non-supersymmetric Yang Mills theory. This point is open here.

%%%%%%%%%%%%%%%%%%%%%%% Fig.7 %%%%%%%%%%%%%%%%%%%%%%%%%%%%
\begin{figure}[htbp]
\includegraphics[width=14cm,height=7cm]{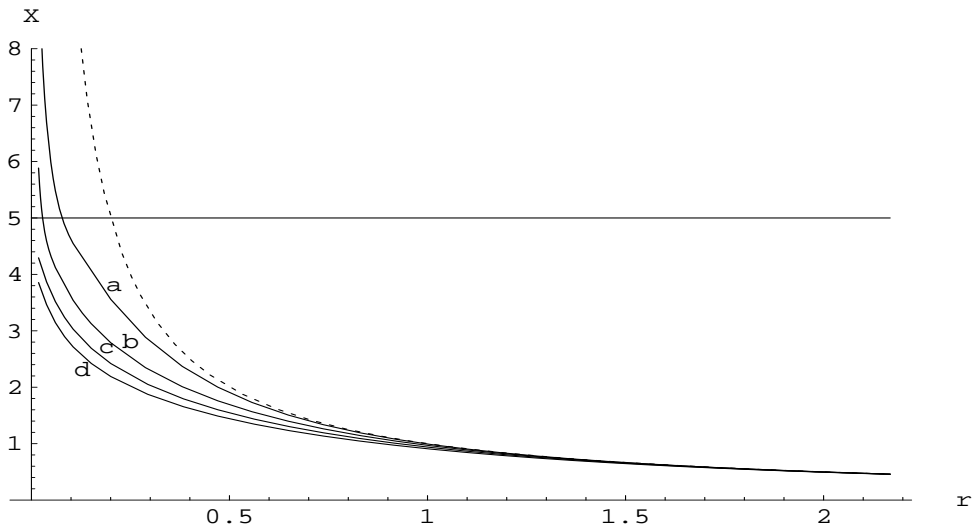}
\caption{\footnotesize{The $\beta$-dependece of the dyonic non-BPS solutions
for $X_m=5, A_2=1$, $\tilde{Q}=1$ with the boundary condition $X(10)=0.1$ 
and $X'(10)=-0.009974$. The curves for (a) $\beta=$0.2, (b) 0.6, (c) 1.0
and (d) 1.4 are shown.
}}
\end{figure}
%%%%%%%%%%%%%%%%%%%%%%%% Fig.7 %%%%%%%%%%%%%%%%%%%%%%%%%%%

Finally, we comment on the solutions of $\beta\neq 0$ 
which are reduced to the
solutions belonging to the group of $(a)\sim (d)$ of Fig.3 for $\beta=0$. The results
are shown in the Fig.7 for various values of $\beta$. 
In this case,
the qualitative behavior of the solutions is the same with the one
of $\beta=0$. But we should notice that there is no bound in principle
at $X=X_m$.

\section{Conclusions}

We have given the solutions of the world-volume action of a D3-brane which
is placed parallel to the background D3-brane(s). The equations are firstly
solved in the flat background, and two kinds of non-BPS solutions 
are shown, $i.e.$ the pure electric and the dyonic solutions.
They smoothly approaches to the BPS solutions 
in a special limit of the parameters. 

Both the BPS and non-BPS solutions are also obtained in the 
D-brane(s) background.
The BPS solutions has the same functional form with the one obtained in the
flat background and it 
can extend to infinity. But it should be bounded at the position of the
opposite brane(s) since
the proper distance in the world-volume of the
brane becomes infinite there. As a result, a finite
energy of this BPS state is obtained. 

On the other hand, both types of non-BPS solutions are affected by the D-brane
background. Especially the electric-type solutions
can not arrive at the 
opposite brane(s) since they are pushed back by the background configurations.
In this sense, this type of non-BPS solutions can not be considered as strings
which connect two branes. One of these non-BPS solutions can be interpreted
as the half-part of the bound state of the brane and anti-brane as in the case
of the flat background. It is observed that the 
distance between the brane and the anti-brane of this bound state becomes
shorter when it nears the background branes.

Although the configurations of the dyonic non-BPS solutions
are also affected by the D-brane background, they
could extend over $X=X_m$ and $X$
becomes infinite at $r=0$. On the other hand, the induced-metric of 
the world-volume action of the D-brane is independent on the solutions, so
the configurations of the solutions are bounded at $X=X_m$ as in the case of 
the BPS solutions. Then we arrive at the conclusion that the
configuration of non-BPS dyonic string connects two D-branes with a finite
distance and with a finite energy. This configuration however receives a
dynamical correction because of no supersymmetry, and this would be related to
the dynamics of the non-supersymmetric Yang -Mills dynamics.

%\section{Discussions}

%%%%%%%%%%%%%%%%%%%%%%%%%%%%%%%%%

%\section{Acknowledgements}
%I am grateful to 
%I would also like to thank 

\vspace {2cm}
%\newpage

\vfill

\begin{thebibliography}{99}

\bibitem{callan} 
C.G.~Callan and J.M.~Maldacena, ``Brane Dynamics from the Born-Infeld Action,'' {{\tt hep-th/9708147}}
\bibitem{gibbons} G.W.~Gibbons, 
``Born-Infeld particles and Dirichlet p-branes,''
Nucl.~Phys. {\bf B514}, 603 (1998), 
{{\tt  hep-th/9709027}}.
\bibitem{hashi} A. Hashimoto, ``The shape of branes pulled by strings,'' 
 ~~Phys. Rev. {\bf D57}, 6441(1998).
\bibitem{bak} D. Bak, J. Lee and H. Min, 
``Dynamics of BPS States in the Dirac-Born-Infeld Theory,'' 
 {{\tt  hep-th/9806149}}.
\bibitem{sav} K.G. Savvidy, ``Brane Death via Born-Infeld String,'' 
 {{\tt  hep-th/9810163}}.
\bibitem{YM} Y.Imamura, 
``Supersymmetries and BPS Configurations on Anti-de Sitter Space,'' 
 {{\tt  hep-th/9807179}}.
\bibitem{CGS} C.G. Callan, A. Guijosa and K.G. Savvidy, 
``Baryons and String Creation from the Fivebrane Worldvolume Action,'' 
 {{\tt  hep-th/9810093}}.
\bibitem{CGST} C.G. Callan, A. Guijosa, K.G. Savvidy and O. Tafjord, 
``Baryons and Flux Tubes in Confining Gauge Thories from Brane Actions,'' 
 {{\tt  hep-th/9902197}}.
\bibitem{gaun} J.P. Gauntlett, C. Koehl, D, Mateos, P.K. Townsend and
 M. Zamaklar, ``Finite energy Dirac-Born-Infeld monopoles and string 
 junctions,''{{\tt hep-th/9903156 }}
\bibitem{CRS} J.M. Camino, A.V. Ramallo and J.M. Sanchez de Santos, 
``Worldvolume Dynamics of D-branes in a D-brane Background,'' 
 {{\tt  hep-th/9905118}}.
\bibitem{Duff} M. J. Duff, R. R. Khuri and J. X. Lu, Phys. Rep., 
{\bf 259}, 213(1995).
\bibitem{Horowitz} G. T. Horowitz and Strominger, Nucl. Phys., 
{\bf A360}, 197(1991).
\bibitem{polchin} J.Polchinski., Phys. Rev. Lett., {\bf 75}, 141(1997),
{{\tt hep-th/9711106 }}
\end{thebibliography}
\end{document}